# Responsivity Calibration of Pyroelectric Terahertz Detectors

Christopher W. Berry[1], Nezih. T. Yardimci[2], Mona Jarrahi[1,2]

Email: mjarrahi@ucla.edu

[1]Electrical Engineering and Computer Science Department, University of Michigan Ann Arbor
[2]Electrical Engineering Department, University of California Los Angeles

There has been a significant advancement in terahertz radiation sources in the past decade, making milliwatt terahertz power levels accessible in both continuous-wave and pulsed operation [1-4]. Such high-power terahertz radiation sources circumvent the need for cryogenic-cooled terahertz detectors such as semiconductor bolometers and necessitate the need for new types of calibrated, room-temperature terahertz detectors. Among various types of room-temperature terahertz detectors, pyroelectric detectors are one of the most widely used detectors, which can offer wide dynamic range, broad detection bandwidth, and high sensitivity levels [5]. In this article, we describe the calibration process of a commercially available pyroelectric detector (Spectrum Detector, Inc, SPI-A-65 THz), which incorporates a 5 mm diameter $LiTaO_3$ detector with an organic terahertz absorber coating to offer a noise equivalent power of 1 $nW/\sqrt{Hz}$.

In order to calibrate the pyroelectric detector, the output power of a plasmonic photomixer mounted on a silicon lens [3] is measured as a function of frequency using both the pyroelectric detector and a calibrated silicon bolometer from Infrared Laboratories. For these measurements, the terahertz power levels are set to be within the dynamic range of both detectors over the whole measurement frequency range. Additionally, the measured output power by the pyroelectric detector is determined by using the pyroelectric detector responsivity value at 94 GHz, $R_{p-vendor}$, that has been provided by the vendor. A tapered copper tube is used to guide the radiated power from the photomixer to the pyroelectric detector. Diameter of the tapered copper tube on the detector side is set to match the diameter of the input aperture of the pyroelectric detector (Fig. 1).

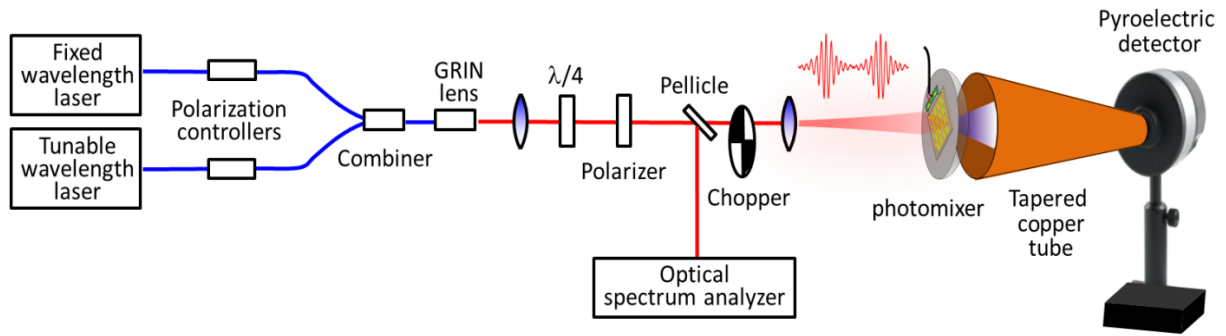

Fig 1. Experimental setup for calibrating the pyroelectric detector responsivity as a function of frequency.

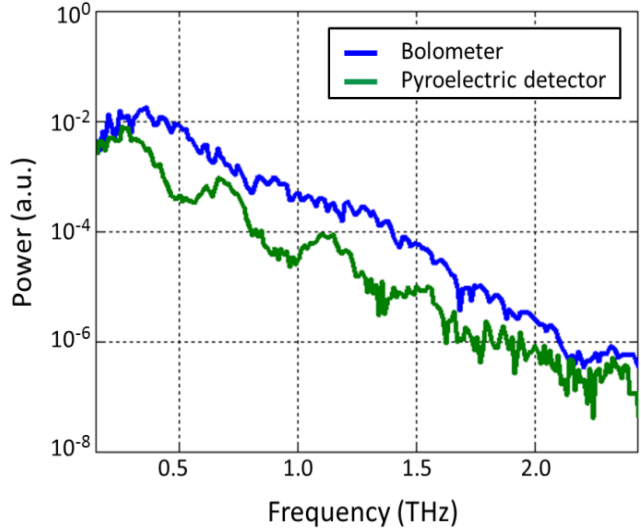

Fig 2. Measured terahertz power by the pyroelectric detector and bolometer.

Figure 2 shows the terahertz power of the photomixer measured by the pyroelectric detector, $P_{pyroelectric}(f)$, and silicon bolometer, $P_{bolometer}(f)$. As expected, the measured power levels by both detectors match at 94 GHz, confirming the accuracy of the pyroelectric detector responsivity data at 94 GHz. Deviations between the measured power levels by the bolometer and pyroelectric detector, which are in the form of Fabry-Perot transmission resonances, are mainly due to the utilized organic coating on the surface of the pyroelectric detector. These deviations are used to calculate the responsivity correction factor of the pyroelectric detector at other frequencies. The calibrated responsivity of the pyroelectric detector as a function of frequency is calculated as

$$R_p(f) = R_{p-vendor} \frac{P_{pyroelectric}(f)}{P_{bolometer}(f)} \qquad (1)$$

The calibrated responsivity of the pyroelectric detector is used to calculate the actual radiation power of various terahertz sources in narrowband/continuous-wave and broadband/pulsed operation [3, 4, 6]. For the specific case of broadband terahertz radiation sources, the calibrated responsivity spectrum is used in combination with the radiation spectrum of the source to measure the actual radiated power. Radiation spectrum of terahertz sources can be measured by various techniques such as time-domain spectroscopy [7, 8] and frequency-domain interferometry [9]. For a broadband terahertz radiation source with a normalized radiation spectrum of $S_{source}(f)$, the voltage reading on the pyroelectric detector and the calculated radiated power according to the vendor-provided responsivity value at 94 GHz are given by

$$V_{pyroelectric} = \int_0^\infty A_0 \cdot S_{source}(f) R_p(f) df = R_{p-vendor} \int_0^\infty A_0 \cdot S_{source}(f) \frac{P_{pyroelectric}(f)}{P_{bolometer}(f)} df \qquad (2)$$

$$P_{calculated} = \frac{V_{pyroelectric}}{R_{p-vendor}} = \int_0^\infty A_0 \cdot S_{source}(f) \frac{P_{pyroelectric}(f)}{P_{bolometer}(f)} df \qquad (3)$$

where $A_0$ is the amplitude of the radiated terahertz power spectrum. Therefore, the actual measured terahertz power by the pyroelectric detector can be calculated by using the calibrated responsivity spectrum:

$$P_{actual} = \int_0^\infty A_0 \cdot S_{source}(f) df = P_{calculated} \frac{\int_0^\infty S_{source}(f) df}{\int_0^\infty S_{source}(f) \frac{P_{pyroelectric}(f)}{P_{bolometer}(f)} df} \qquad (4)$$

It should be noted that the use of the tapered copper tube for guiding the radiated terahertz beam into the input aperture of the pyroelectric detector does not impact the accuracy of the presented calibration technique as far as the same copper tube is used as a part of the pyroelectric detector package for terahertz power measurements. As such, a similar approach can be used for calibrating other types of terahertz detectors and the calibration data can be used for various terahertz power measurements in narrowband/continuous-wave and broadband/pulsed operation [3, 4, 6].

A point of caution when using thermal detectors (e.g. pyroelectric detector and bolometer) for terahertz power measurements is to make sure that all the detected power is the contribution of terahertz radiation, not the induced heat in the experimental setup. This can be very important when measuring the radiation power of photoconductive terahertz sources and photomixers at high optical pump power and bias voltage levels. In order to quantify the potential contribution of thermal power to the measured terahertz power levels by a thermal detector, power measurements should be repeated under a single-frequency continuous-wave optical pump beam at the same bias voltage, pump wavelength and pump power levels that the terahertz power measurements are taken. Comparisons between the measured power levels under a single-frequency continuous-wave optical pump beam and the measured terahertz power levels under a heterodyning continuous-wave or femtosecond pulsed optical pump beam will determine the potential contribution of thermal power to the measured terahertz power levels.

**References**

[1] E. Peytavit, S. Lepilliet, F. Hindle, C. Coinon, T. Akalin, G. Ducournau, G. Mouret, and J.-F. Lampin, "Milliwatt-level output power in the sub-terahertz range generated by photomixing in a GaAs photoconductor," Appl. Phys. Lett. 99, 223508, 2011

[2] C. W. Berry, M. R. Hashemi, and M. Jarrahi, "Generation of High Power Pulsed Terahertz Radiation using a Plasmonic Photoconductive Emitter Array with Logarithmic Spiral Antennas," Appl. Phys. Lett. 104, 081122, 2014

[3] C. W. Berry, M. R. Hashemi, S. Preu, H. Lu, A. C. Gossard, and M. Jarrahi, "High Power Terahertz Generation Using 1550 nm Plasmonic Photomixers," Appl. Phys. Lett. 105, 011121, 2014

[4] N. T. Yardimci, S.-H. Yang, C. W. Berry, and M. Jarrahi, "Plasmonics Enhanced Terahertz Radiation from Large Area Photoconductive Emitters," Proc. IEEE Photon. Conf., San Diego, CA, October 12-16, 2014

[5] https://www.gentec-eo.com/products/thz-detectors

[6] S.-H. Yang, M. R. Hashemi, C. W. Berry, and M. Jarrahi, "7.5% Optical-to-Terahertz Conversion Efficiency Offered by Photoconductive Emitters with Three-Dimensional Plasmonic Contact Electrodes," IEEE Trans. THz Sci. Technol. 4, 575-581, 2014

[7] Y. Cai, I. Brener, J. Lopata, J. Wynn, L. Pfeiffer, J. B. Stark, Q. Wu, X. C. Zhang, and J. F. Federici, "Coherent terahertz radiation detection: direct comparison between free-space electro-optic sampling and antenna detection," Appl. Phys. Lett. 73, 444-446, 1998

[8] C. W. Berry, M. R. Hashemi, M. Unlu, M. Jarrahi, "Design, Fabrication, and Experimental Characterization of Plasmonic Photoconductive Terahertz Emitters," J. Vis. Exp. 77, e50517, 2013

[9] M. R. Hashemi, C. W. Berry, E. Merced, N. Sepulveda, and M. Jarrahi, "Direct Measurement of Vanadium Dioxide Dielectric Properties in W-band," J. Infrared Millim. THz Waves 35, 486-492, 2014